\documentclass[aps, pra, reprint, superscriptaddress,longbibliography]{revtex4-2}
\usepackage{lineno,hyperref}
\usepackage{eurosym}
\usepackage{amsmath,amssymb,graphicx}
\usepackage{color}
\usepackage{float}
\usepackage{appendix} 
\usepackage[normalem]{ulem}

\modulolinenumbers[5]
\newcommand{\bitem}{\begin{itemize}}
\newcommand{\fitem}{\end{itemize}}
\newcommand{\beq}{\begin{equation}}
\newcommand{\eeq}{\end{equation}}
\newcommand{\beqa}{\begin{eqnarray}}
\newcommand{\eeqa}{\end{eqnarray}}

\begin{document}

\title{\textbf{Superradiant Phase Transition and Statistical Properties in the Dicke-Stark Model} 
}%
\author{Weilin Wang}
\affiliation{Department of Mathematics and Physics, North China Electric Power University, Huadian Road, Baoding 071000, China}
\author{Ronghai Liu}
\affiliation{Electric Power Research Institute of Yunnan Power Grid Co., Ltd., Kunming 650011, China}
\author{Fangcheng Qiu}
\affiliation{Electric Power Research Institute of Yunnan Power Grid Co., Ltd., Kunming 650011, China}
\author{Mingshu Zhao}
\affiliation{Department of Mathematics and Physics, North China Electric Power University, Huadian Road, Baoding 071000, China}
\author{Jinying Ma}
\affiliation{Department of Mathematics and Physics, North China Electric Power University, Huadian Road, Baoding 071000, China}
\author{Zhanyuan Yan}
\email{Contact author: yanzhanyuan@ncepu.edu.cn}
\affiliation{Department of Mathematics and Physics, North China Electric Power University, Huadian Road, Baoding 071000, China}
\affiliation{Hebei Key Laboratory of Physics and Energy Technology, North China Electric Power University, Baoding 071003, China}
\date{\today}
          
\begin{abstract}

In this study, we investigate the statistical properties and dynamical behaviors of the finite-size Dicke-Stark model. We numerically obtain the energy spectrum and eigenstates within the extended coherent state space, and subsequently utilize the quantum dressed master equation to describe the open system dynamics under strong coupling. Under thermal equilibrium conditions, analyzes of the negativity, zero-time-delay two-photon correlation function, and atom-spin squeezing parameters reveal that as the coupling strength increases, the light field undergoes a transition from photon bunching to anti-bunching and then back to bunching. The Stark field can modulate both the extrema of the two-photon correlation function and their corresponding coupling strengths. At low temperatures, the system exhibits entanglement and spin squeezing. As the temperature increases, the entanglement gradually diminishes, while strong coupling facilitates the preservation of the entanglement. The atom-spin squeezing is highly sensitive to thermal fluctuations and vanishes rapidly with increasing temperature. Furthermore, we illustrate the time-evolution properties of the thermodynamic non-equilibrium of the system from a low-temperature state to a high-temperature state. Our results demonstrate that an appropriately tuned Stark interaction can effectively prolong the survival time of quantum entanglement against thermal noise. This work contributes to the fundamental understanding of quantum phenomena in Dicke-Stark systems.

\end{abstract}
\maketitle 
\section{INTRODUCTION}
The light-matter interaction is a significant area of research, both theoretically and experimentally, spanning various fields, including condensed matter physics, quantum information, and quantum optics~\cite{cohen1993atom}. 
In 1936, Rabi proposed the foundational theoretical model to describe this interaction~\cite{rabi1936process}: the coupling between a single-mode quantum optical field and a single two-level atom represents the simplest coupled system in quantum optics. Despite its conceptual simplicity, the Rabi model is difficult to solve analytically. Limited by the experimental conditions at that time, the coupling strength between light and the two-level system was relatively weak, allowing the counter-rotating wave terms in the Rabi model to be neglected (the rotating-wave approximation, RWA), which simplified the Rabi model into the Jaynes-Cummings (J-C) model~\cite{jaynes2005comparison}. Owing to its analytical solvability, the J-C model was widely applied in research across related fields for a considerable period thereafter. With the advancement of experimental techniques, around 2010, experimental platforms such as cavity QED~\cite{thompson1992observation,de2018breakdown},  superconducting quantum circuits~\cite{niemczyk2010circuit,yoshihara2017superconducting,forn2010observation,2017Multi,chen2017single,yoshihara2018inversion}, and optomechanical systems~\cite{benz2016single,pirkkalainen2015cavity} successively achieved strong, ultra-strong and deep-strong coupling between light and the system. These experiments confirmed the limitations of the J-C model, highlighting the need to analytically solve the Rabi model ~\cite{cohen1993atom}. 
In 2011, Braak achieved a theoretical breakthrough by analytically resolving the quantum Rabi model by exploiting its symmetry in Bargmann space~\cite{braak2011integrability}. In 2012, Chen proposed the Bogoliubov Operator Approach (BOA) method~\cite{chen2012exact}, which reproduced the analytical solution of the quantum Rabi model, providing a more precise and concise physical interpretation.
In 2013, Zhong derived the exact solution of the Rabi model expressed in terms of the confluent Heun function~\cite{zhong2014analytical}.
Subsequently, researchers have explored more complex models of light-matter interactions, including two-photon Rabi model~\cite{maciejewski2015comment,zhang20172,cui2017exact,duan2016two}, two-mode Rabi model~\cite{duan2015solution,yan2022analytical,liu2023spectral}, anisotropic Rabi model~\cite{tomka2014exceptional,shen2014ground,zhang2015analytical,travvenec2012solvability}, the Rabi-Stark model~\cite{xie2020first,xie2019quantum,li2020two}, and the Dicke model~\cite{braak2013solution,he2014exact}.

The challenge in resolving the quantum Rabi model originates from the presence of the counter-rotating wave term. This term leads to the states of the system being the superpositions of all Fock basis states. As a result, the Hamiltonian of the system is presented as an irreducible infinite-dimensional matrix. Theoretically, the larger the matrix dimension, the more accurate the calculation results, which imposes rather high requirements on the data storage and computing capabilities of computers. 

In the case of the more intricate Dicke model, which describes the interaction between $N$ two-level atoms and a single-mode light field, the dimensionality of the system space is obtained from the direct product of the angular momentum space and the Fock state space (photon number space). Obviously, when $N \to \infty $, it is impossible to numerically solve the Dicke model. However, collective symmetry enables analytical derivation of key properties like the ground state and superradiant phase transition (SPT) ~\cite{chen2008numerically,emary2003quantum}.
For the finite-size Dicke model consisting of a finite number of atoms, due to the lack of collective symmetry, numerical diagonalization of the full energy spectrum and eigenstates imposes substantial computational burdens.
The extended coherent state space is constructed by applying a displacement operator to the Fock state space. Notably, it has been found that within this space, only a finite number of basis vectors is required to solve the eigen-equation of the Dicke model with sufficient accuracy~\cite{zhang2010quantum}. 
Owing to the implementation of numerical solutions for the Dicke model, it has been applied to the design of quantum heat engines and quantum batteries~\cite{xu2024universal,dou2022extended}.
Researchers are exploring strategies to enhance their performance, where temperature emerges as a critical factor: thermal environmental noise induces relaxation and decoherence effects that limit overall efficiency \cite{ji2022spin, zheng2016occurrence, li2018efficient,  rasola2024autonomous}. This necessitates investigating the Dicke model in open quantum systems. In References \cite{le2016fate,Xu2020Dicke,xu2024exploring}, under weak coupling between the Dicke system and an Ohmic heat bath, the dressed master equation was derived using the Born-Markov approximation. This equation accurately describes the non-unitary evolution of two-level systems in thermal environments. Importantly, the dressed master equation is applicable to strongly coupled Dicke systems.

In the process of quantum state preparation and light-field pattern reconstruction, a nonlinear Stark interaction is introduced between the optical field and two-level systems~\cite{pellizzari1994preparation,santos2001conditional,grimsmo2013cavity,grimsmo2013strong}.
Previous studies have examined Stark effects on spectral structures by incorporating these interactions into the Rabi model~\cite{xie2020first,xie2019quantum,li2020two}. In the Dicke model, a nonlinear Stark field is introduced, which is called the Dicke-Stark (DS) model~\cite{mu2020dicke}. 
In this paper, the dressed master equation of the DS system will be calculated within the extended coherent state space.  
Subsequently, we will further investigate the dynamical properties of the Dicke model. As presented in the following sections, we analytically derive the critical point of the SPT  both at zero and finite temperatures of the infinite-size DS model using mean-field theory. Additionally, we numerically confirm the existence of SPT in the finite-size DS model, demonstrating that the critical point can be continuously tuned by the Stark field strength. This tunability offers a valuable mechanism for designing and improving the performance of quantum heat engines \cite{xu2024exploring}.
Furthermore, we study the quantum statistical properties of the DS system, such as the two-photon correlation function, negativity, and spin squeezing parameters of the system's equilibrium state, which are employed to discuss the entanglement of the system's equilibrium state, the correlation of the light field, and the squeezing properties of the atomic state, respectively. 
Several findings are particularly interesting: The two-photon correlation function identifies a dynamical transition process of the photon state, evolving from a thermal state to quantum anti-bunching, then to bunching, and finally back to a thermal state, with the entire process modulated by Stark field strength. The negativity reveals the sustainability of entanglement in the system state, which is modulated by Stark field strength, degrades as temperature increases, and exhibits low-temperature robustness under strong coupling. Atomic state squeezing is enhanced by negative Stark field at low temperatures but shows extreme thermal sensitivity.

The paper is organized as follows. In Section II, we introduce the quantum DS model, along with its numerical solution method and the construction of the dressed master equation for thermal equilibrium.
In Section III, we performed analytical solutions for the superradiant phase transition in the ground state at both zero and finite temperatures.
In Section IV, the ground-state average photon number is calculated, which reproduces the quantum phase transition.
Moreover, the time-dependent  average photon number will be analyzed.
In Section V, the thermal equilibrium properties of the open finite-size Dicke-Stark model were investigated, characterized by the zero-time delay two-photon correlation function, negativity, and spin squeezing. In Section VI, we summarize the main findings.

\section{Dicke-Stark Model and Solution Methods}
\subsection{\label{sec:level2-2}Dicke-Stark Model}
The Dicke-Stark model describes a quantum system composed of $N$ two-level qubits interacting with a single-mode bosonic field. The Hamiltonian of the system can be expressed as($ \hbar = 1$\ and $ k _ { B } = 1$)~\cite{garraway2011dicke,gopalakrishnan2011frustration,
bastidas2012nonequilibrium,abdel2017quantum,mu2020dicke}
:
\begin{equation}~\label{H_D}
{\hat H_{DS}} = \omega {\hat a^ \dagger  }\hat a + \Delta {\hat J_z} + \frac{{2\lambda }}{{\sqrt N }}({\hat a^ \dagger  } + \hat a){\hat J_x} + \frac{U}{N}{\hat a^ \dagger  }\hat a{\hat J_z},
\end{equation}
where ${\hat J_x} $ and $\hat { J } _ { z }$ represent the collective
spin operators, composed of $\hat { J } _ { \alpha } = \sum _ { i=1 } ^ { N } \frac{1}{2}\hat { \sigma } _ { \alpha } ^ { i }$ , with $ \hat { \sigma } _ { \alpha } ( \alpha = x, y, z)$ as the Pauli operators. ${\hat a^ \dagger  }$ and ${\hat a }$ denote the creation and annihilation operators of the bosonic field, $\Delta $ and $\omega $ represent the frequency of the qubits and the single bosonic mode, respectively, and $\lambda$ is the qubit-boson coupling strength. $\ U$ represents the strength of the nonlinear Stark interaction. In the following section, we set the frequency of bosonic $\omega$ as the energy unit for simplicity.

Due to the lack of complete energy spectral and eigenstate information for the DS model in the thermodynamic limit ($N \to \infty$), numerically computing the energy spectra and eigenstates of the finite-size DS model within the bosonic field's Fock space requires a large photon-number cutoff to ensure computational accuracy. This necessitates solving a Hamiltonian matrix of enormous dimension, posing significant challenges to computer memory and computational time.

\subsection{\label{sec:level2-3}The extended coherent state approach}
An extended coherent bosonic state approach is proposed to accurately calculate the energy spectrum and eigenstates of the finite-size Dicke model with a small photon number cutoff~\cite{chen2008numerically}. This method is also applicable to the finite-size DS model. Before the introduction of the extended coherent bosonic state approach, we rotate the collective spin operators with $\pi /2$ along $\hat{J}_{y}$ by ${\hat H} = \exp(i\pi {\hat J_y}/2){\hat H_{DS}}\exp( - i\pi {\hat J_y}/2)$, resulting in
\begin{equation}~\label{H}
\hat H = \omega {\hat a^\dagger }\hat a - (\frac{\Delta }{2} + \frac{U}{{2N}}{\hat a^\dagger }\hat a)({\hat J_ + } + {\hat J_ - }) + \frac{{2\lambda }}{{\sqrt N }}({\hat a^\dagger } + \hat a){\hat J_z}.
\end{equation}
The eigen-equation can be written in terms of coherent state methods (see Appendix \ref{FA}):
\begin{equation}~\label{equ:E_n}
\begin{array}{l}
\omega C_{n,l}^n(k - g_n^2)\\
 - \left[ {\frac{\Delta }{2} + \frac{U}{{2N}}(k + g_{_{n - 1}}^2)} \right]C_{_{n - 1},l}^nj_{_{n - 1}}^ + \sum\limits_k {{}_{{A_n}}\left\langle l | k \right\rangle }_{{A_{n - 1}}} \\
 + \left[ {\frac{\Delta }{2} + \frac{U}{{2N}}(k + g_{_{n + 1}}^2)} \right]C_{_{n + 1},l}^nj_{_{n + 1}}^ - \sum\limits_k {{}_{{A_n}}\left\langle l|k \right\rangle }_{{A_{n + 1}}} \\
 + \frac{U}{{2N}}C_{n - 1,k}^n{g_{n - 1}}\sqrt {k + 1} j_{n - 1}^ + \sum\limits_k {{}_{{A_n}}\left\langle l|{k + 1} \right\rangle }_{{A_{n - 1}}} \\
 + \frac{U}{{2N}}C_{n - 1,k}^n{g_{n - 1}}\sqrt k j_{n - 1}^ + \sum\limits_k {{}_{{A_n}}\left\langle l | {k - 1} \right\rangle }_{{A_{n - 1}}} \\
 + \frac{U}{{2N}}C_{n + 1,k}^n{g_{n + 1}}\sqrt {k + 1} j_{n + 1}^ - \sum\limits_k {{}_{{A_n}}\left\langle l | {k + 1} \right\rangle }_{{A_{n + 1}}} \\
 + \frac{U}{{2N}}C_{n + 1,k}^n{g_{n + 1}}\sqrt k j_{n + 1}^ - \sum\limits_k {{}_{{A_n}}\left\langle l | {k - 1} \right\rangle }_{{A_{n + 1}}} \\
 = {E_n}C_{n,l}^n,
\end{array}
\end{equation}
in which, the inner products are
\begin{equation}
    _{A_n}{\langle l|k\rangle _{{A_{n - 1}}}} = {( - 1)^l}{D_{l,k}},_{A_n}{\langle l|k\rangle _{{A_{n + 1}}}} = {( - 1)^k}{D_{l,k}},\notag
\end{equation}
\begin{equation}
{D_{l,k}} = {e^{ - {G^2}/2}}\sum\limits_{r = 0}^{\min (l,k)} {\frac{{{{( - 1)}^{ - r}}\sqrt {l!k!} {G^{l + k - 2r}}}}{{(l - r)!(k - r)!r!}}} , G = \frac{{2\lambda }}{{\omega \sqrt N }}. \notag
\end{equation}
Then, the energy spectrum and eigenstates of the DS model can be numerically calculated using Eq. (\ref{equ:E_n}). We systematically validated the applicability and convergence of the DCS and DFS methods in the DS model, as detailed in Appendix \ref{FB}. In the following work, we select the truncation number $K_{tr} = 50$, which is sufficient to yield convergent state energies and wavefunctions with calculation errors less than ${10^{-6}}$.

\begin{figure*}
\includegraphics[width=0.99\textwidth]{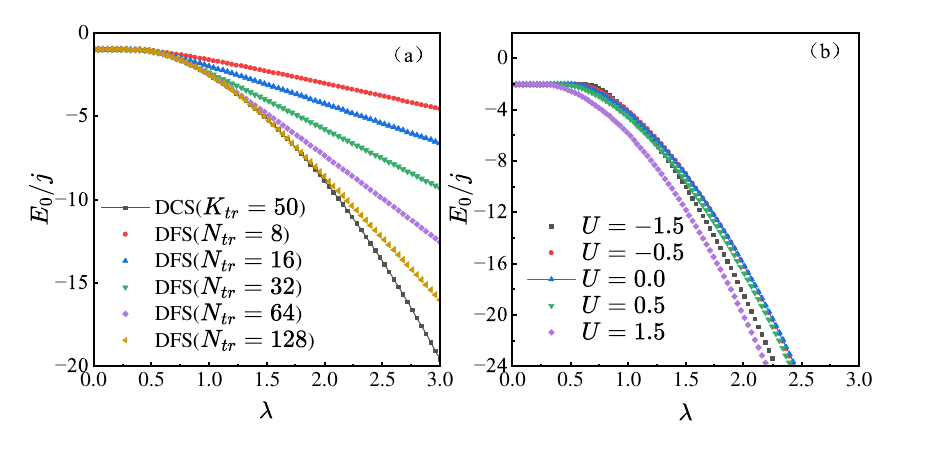}[t]
\caption{\label{DS_Ground_state}Ground-state energy spectrum of DS model as a function  increasing coupling strength $\lambda$. (a). A comparison of ground-state spectrum calculated by DCS method  with photon truncation number $K_{tr}=50$ and by the DFS method with photon truncation numbers $N_{tr}= 8, 16, 32, 64, 128$ at fixed Stark parameter $U=1.0$. (b). Demonstrating the effects of the stark strength $U = 1.5,0.5,0,-0.5, -1.5$ on the ground-state energy spectrum, simulated using the DCS method with $K_{tr}=50$, $N=32$, $\Delta=1$.}
\end{figure*}

To more clearly illustrate the differences in the ground-state energy spectra between diagonalization in coherent states (DCS) and Fock states (DFS), we compute the ground-state energy spectra using the DFS method at parameters \(\Delta = 1.0\) and \(U = 1.0\), with photon number truncations of \(N_{tr} = 8\), \(16\), \(32\), \(64\), and \(128\). Simultaneously, we use the DCS method with a truncation of \(K_{tr} = 50\). These results are shown in Fig.~\ref{DS_Ground_state}(a). As \(N_{tr}\) increases in the DFS method, the calculated ground-state energy spectra gradually approach those obtained by the DCS method. 

Furthermore, Fig.~\ref{DS_Ground_state}(b) presents the ground-state energy spectra for the DS model at $U=1.5,0.5$ and $U=-1.5,-0.5$, alongside the Dicke model at $U=0$. The figures clearly demonstrate that the ground-state energy of the Dicke model is suppressed due to the effects of the nonlinear Stark interaction.


\subsection{\label{sec:level2-4}Quantum dressed master equation}
When the interaction between the optical field and the atoms reaches the strong-coupling regime, the system state should be described using dressed states. Correspondingly, the evolution of the state in the presence of an external environment follows a dressed master equation. In the case of the strong-coupling DS model, the Stark effect only modifies the energy level structure of the system without altering its symmetry. Therefore, following the method described in reference~\cite{mccauley2020accurate}, the introduction of the $U$-term does not affect the derivation of the master equation, and the dressed-state master equations of the Dicke-Stark model and the Dicke model have the same form.

A high-frequency cutoff and a flat spectral density ensure the positivity and Markovian nature of the dynamics. Under these conditions, the system's quantum master equation read as~\cite{gorini1976completely,lindblad1976generators,weiss2012quantum,breuer2002theory}
\begin{equation}~\label{eq:master_equation}
\begin{array}{cc}
\frac{d}{{dt}}{{\hat \rho }_s} =  - i\left[ {{{\hat H}_{DS}},{{\hat \rho }_s}} \right] + \sum\limits_{u;k < j}\{{\Gamma _u^{jk}{n_u}({\Delta _{jk}})D\left[ {|{\phi _j}\rangle \langle{\phi _k}|,{{\hat \rho }_s}} \right]}  \\
 + \Gamma _u^{jk}\left[ {1 + {n_u}({\Delta _{jk}})} \right]D\left[ {|{\phi _k}\rangle \langle {\phi _j}|,{{\hat \rho }_s}} \right]\} ,
\end{array}
\end{equation}
where $D\left[ {\hat O,\hat \rho } \right] = \frac{1}{2}\left[ {2\hat O{{\hat O}^ \dagger } - \hat \rho {{\hat O}^ \dagger }\hat O - {{\hat O}^ \dagger }\hat O\hat \rho } \right]$. Dissipation rates $\Gamma _u^{jk} = {\gamma _u}({\Delta _{jk}})|S_u^{jk}{|^2}$  depend on the spectral function ${\gamma _u}({\Delta _{jk}})$ and the transition coefficients: $S_a^{jk} = \langle {\phi _j}|({\hat a^\dagger} + \hat a)|{\phi _k}\rangle $, $S_{\sigma  - }^{jk} = \langle {\phi _j}|(\hat \sigma  + {\hat \sigma _ - })|{\phi _k}\rangle $. For the Ohmic case, ${\gamma _u}({\Delta _{jk}}) = \pi \alpha {\Delta _{jk}}exp( - |{\Delta _{jk}}|/{\omega _c})$, where $\alpha $ is the coupling strength and ${\omega _{\rm{c}}}$ is the cutoff frequency, throughout all numerical simulations performed, we consider $\alpha  = 0.001\omega $ and ${\omega _c} = 10\omega $. The Bose-Einstein distribution ${n_u}({\Delta _{jk}},{T_u}) = 1/\left[ {exp({\Delta _{jk}}/{T_u}) - 1} \right]$ accounts for thermal effects. The steady-state solution of Eq. (\ref{eq:master_equation}) yields the density matrix of the canonical ensemble, as confirmed by straightforward numerical simulations, given by:
\begin{equation}~\label{equ:density}
    {\hat \rho _{ss}} = \sum\limits_n {\frac{{{e^{ - {E_n}/T}}}}{Z}} |{\phi _n}\rangle \langle {\phi _n}|,
\end{equation}
where $Z = \sum\limits_n {{e^{ - {E_n}/T}}} $ is the partition function.

\begin{figure*}
\includegraphics[width=0.99\textwidth]{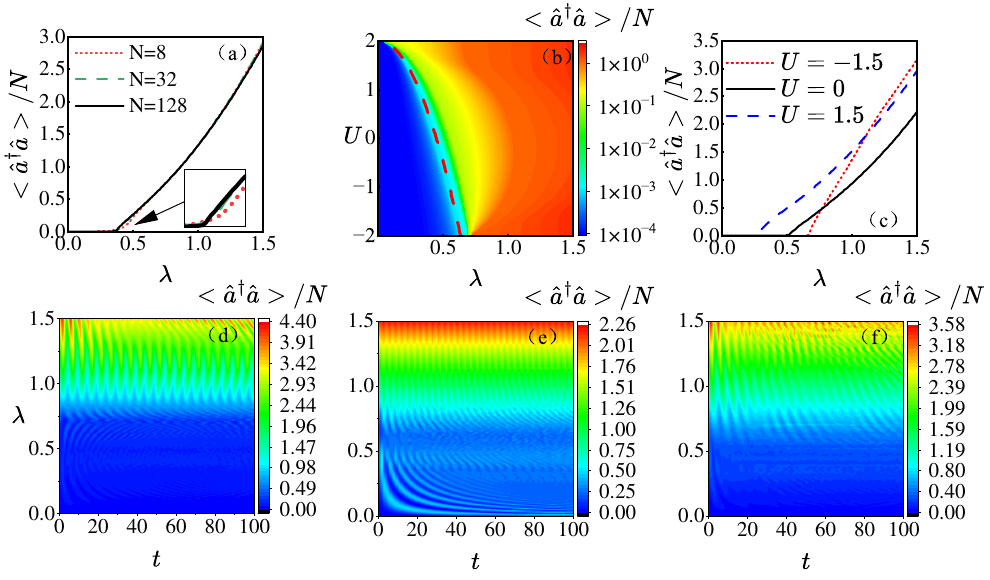}
\caption{~\label{DS_photon_number}
Variations in the average photon number of the ground state in the isolated finite-size Dicke-Stark model:
(a) The variation of the average photon number with coupling strength for different atomic numbers \(N=8, 32,128\) at \(U=1\).
(b) Phase diagram of the average photon number in the plane of coupling strengths and Stark field strengths \(\lambda-U\) with atomic number \(N=128\).
(c) The variation of the average photon number with coupling strength for different Stark field strengths $U=-1.5, 0, 1.5$ at \(N=128\).
(d)-(f) Evolution of average photon number as a function of coupling strength for different Stark field strengths $U=-1.5, 0, 1.5$ with atomic number  $N=16$. Other parameters: \(K_{tr}=50, \Delta=1\).} 
\end{figure*}

\section{\label{sec:SPT} Theoretical solution of  SPT in infinitely sized Dicke-stark model}

To provide a theoretical benchmark for the finite-size numerical results in Secs. IV and V, we summarize the critical conditions for the superradiant phase transition (SPT) in the thermodynamic limit (\(N \to \infty\)). The detailed mean-field derivations are provided in Appendix \ref{FC}.

For the isolated DS system at zero temperature, the critical coupling strength \(\lambda_c\) is derived using the Holstein-Primakoff transformation (see Appendix \ref{FC1}):
\begin{equation}
\lambda_c = \frac{1}{2}\sqrt{\Delta\left(\omega - \frac{U}{2}\right)}.
\label{equ:lambda_c}
\end{equation}
Extending this to finite temperatures \(T\), the critical point is determined by the instability of the normal phase in the free energy landscape (see Appendix \ref{FC2}), yielding:
\begin{equation}
\lambda_c(T) = \frac{1}{2} \sqrt{\Delta \left[ \frac{\omega}{\tanh\left(\frac{\Delta}{2k_B T}\right)} - \frac{U}{2} \right]}.
\label{equ:lambda_c(T)}
\end{equation}
In the low-temperature limit (\(T \to 0\)), Eq. (\ref{equ:lambda_c(T)}) naturally recovers the zero-temperature result of Eq. (\ref{equ:lambda_c}).

We remark that the critical conditions derived above are consistent with recent developments in generalized Dicke models. Specifically, Eqs. (\ref{equ:lambda_c}) and (\ref{equ:lambda_c(T)}) correspond exactly to the isotropic (\(\tau = 1\)) and \(A^2\)-term-free (\(\kappa = 0\)) limits of the critical conditions reported for the anisotropic Dicke-Stark model \cite{chen2024phase}.

While the phase boundaries are analytically encompassed by the generalized framework, the specific impact of the Stark term on the quantum statistical properties—such as photon correlation and spin squeezing dynamics discussed in Sec. V—remains an open question, which we address numerically in the following sections.

\section{SPT in the isolated finite-size DS model}
To discuss the SPT in the finite-size isolated DS model, we calculate the average number of photons \(\langle \hat{a}^\dagger \hat{a} \rangle / N\) of each atom in the ground state. In the extended coherent state representation, the matrix elements of the photon number operator $\hat N=\hat{a}^\dagger \hat{a}$ are denoted as:
\begin{equation}~\label{photon_number}
\begin{array}{l}
    _{A_{{m^\prime }}} < {k^\prime }| < j,{m^\prime }|{\hat a^\dagger }\hat a |j,m > |k{ > _{{A_m}}} =\\
    \left[ {(k + g_m^2){\delta _{{k^\prime }k}} - {g_m}(\sqrt {k + 1} {\delta _{{k^\prime }k + 1}} + \sqrt k {\delta _{{k^\prime }k - 1}})} \right]{\delta _{{m^\prime }m}},
\end{array}
\end{equation}
Figure~\ref{DS_photon_number} shows the influence of coupling strength $\lambda$, Stark strength $U$, and the sizes of the DS model on the average photon number in the ground state, as well as the phase diagram of its evolution. Panel (a) depicts the variation of the average number of photons with coupling strength for different sizes of the DS model (\(N = 8\), 32, and 128) with \(U = 1.0\). It can be observed that the average photon number exhibits an abrupt increase near the critical point \(\lambda_c\), indicating the occurrence of a SPT. The enlarged image in panel (a) demonstrates that for $N = 8,$ the increase in the average photon number is relatively gradual, making it difficult to identify the precise transition point. This is because the SPT is caused by the collective symmetry of the infinite-size DS model. When the size of the DS model is too small, the collective symmetry is lost, resulting in an indistinct SPT point. At \(N = 128\), the SPT point \(\lambda_c = 0.35\) can be distinguished, which is basically consistent with the calculation result of Eq.~(\ref{equ:lambda_c}).
Panel (b) shows phase diagrams of average photon number in the $\lambda-U$ planes of the finite-size DS model. We calculate the
average photon number $\langle \hat{a}^\dagger \hat{a} \rangle/N$ by exact numerical diagonalization of Hamiltonian in Eq.~(\ref{photon_number}) with
$N = 128$ atoms. $\langle \hat{a}^\dagger \hat{a} \rangle/N$ is almost zero below the critical line (red dashed line) and then smoothly increases in the superradiant phase regime. The analytical coupling strength $\lambda_c$ dependent on $U$ in Eq.~(\ref{equ:lambda_c}) fits well with the critical phase boundary in panel (b). Although the nonlinear Stark coupling is not necessary for the occurrence of the SPT, it can effectively reduce the critical coupling point in practical experiments. This can be clearly seen in panel (c), where the position of the SPT point significantly shifts for different Stark field strengths $\ U=-1.5, 0, 1.5$. As $U$ increases, the corresponding critical coupling strength $\lambda_c$ decreases. The location of the phase transition point appears to coincide with the result of Eq.~(\ref{equ:lambda_c}), which is about $\lambda=0.25$ when $\ U=1.5$, while it is $\lambda=0.66$ when $\ U=-1.5$ and exactly $\lambda=0.5$ when $\ U=0$. $\lambda_c$ decreases to zero as $U$ approaches 2.0. 


In the following, we discuss the evolution properties of the isolated DS model by calculating the average number of photons $ < {\hat a^\dagger }\hat a > /N$ with respect to time. 
Assume the initial state $|\psi (0)\rangle  = |j, - j\rangle |0{\rangle _a}$, the wave function of the system at time $t$ is given by
\begin{align}
    |\psi (t)\rangle  &= {e^{ -iHt}}|\psi (0)\rangle  \\&=\sum\limits_n {d_n^0} {e^{ - i{E_n}t}}\sum\limits_n {C_{m,k}^n} |j,m > |k{\rangle _{{A_m}}},
\end{align}
where $d_n^0 = \langle {\psi _n}|\psi (0)\rangle  = \sum\limits_k {C_{j,k}^n} \frac{{{{( - 1)}^k}}}{{\sqrt {k!} }}{(\frac{{\lambda \sqrt N }}{\omega })^k}{e^{ - \frac{{{\lambda ^2}N}}{{2{n^2}}}}}$.
Within the extended coherent state space, the expectation value of the photon number operator ${\hat a^\dagger }\hat a $ is expressed as
\begin{equation}
 \begin{array}{l}
\langle {\hat a^\dagger }\hat a\rangle  =\langle \psi (t)|{\hat a^\dagger }\hat a|\psi (t)\rangle=\\ 
\sum\limits_{l,n} {\left\{ {d_l^{0*}d_n^0{e^{i({E_l} 
- {E_n})t}}} \right.} \sum\limits_{m,k} [C_{m,k}^{l*}{C_{m,k}^n}(k + g_m^2) 
\\- {g_m}(C_{m,k + 1}^{l*}C_{m,k}^n\sqrt {k + 1}  + C_{m,k - 1}^lC_{m,k}^n\sqrt k )]\} .
\end{array}  
\end{equation}

Due to computational complexity, we adopt a system size of $N=16$ atoms in the DS model to investigate the time evolution of the average photon number.
Figures~\ref{DS_photon_number}(d)-(f) display the phase diagrams of 
$ < {\hat a^\dagger }\hat a > /N$  in the $\lambda-t$ plane for \(U = -1.5,0, \) and \(1.5 \), respectively.
It can be observed that, as a result of a small atomic number \(N = 16\) used in the calculations, the SPT point becomes indistinct, leading to the formation of a banded region near $\lambda = 0.5$ where the color gradient shifts from blue to green in panels (d)-(f). Compared with the case of \(U = 0\), it is found that when \(U > 0\), the overall characteristics of the phase diagram shift towards smaller coupling strengths, while when \(U < 0\), the shift is towards larger coupling strengths. For $U<0$, the average photon number increases sharply with growing $\lambda$ (evident from the steep color gradients in panels (d)-(f)), while for \(U \geq 0\), the increases are more gradual. These behaviors align with the observations in panel (c).
We further found that, for non-zero \(U\) (\(U \neq 0\)), the time evolution of the average photon number is smoother, whereas for \(U = 0\), clear oscillatory evolution is observed in the weak coupling region.

\section{\label{sec:properties}Statistical properties of open finite-size DS model}

In this section, utilizing the energy spectrum and the corresponding states of the open DS system calculated in the coherent-state phase space above, we will discuss quantum correlation properties of the DS model:
the two-photon correlation function, which describes the statistical properties of the photon-field; the negativity, which quantifies the thermal state entanglement; and the squeezing parameter, which quantifies the degree of atom-spin-state squeezing.

\subsection{\label{sec:properties1}Statistical properties of the thermodynamic equilibrium states }

First, we investigate the thermodynamic equilibrium properties of the DS model governed by the dressed master equation [see Eq. (\ref{equ:density})].

\subsubsection{\label{sec:correlation} Two-photon correlation function}

\begin{figure*}
\includegraphics{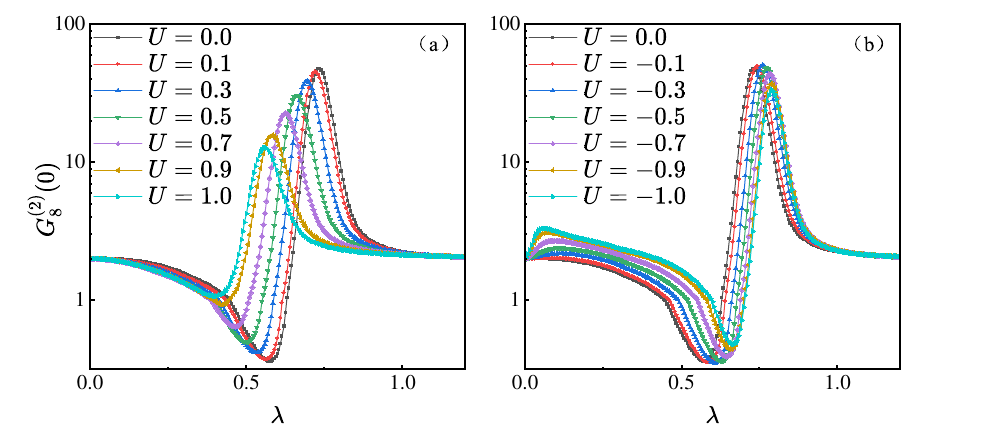}
\caption{The zero-time delay two-photon correlation function ${G^{(2)}}(0)$ as a function of the coupling strength $\lambda$. ${G^{(2)}}(0)$. Panel (a) demonstrates the variation for positive Stark field strengths, while (b) illustrates the case for negative Stark field strengths. In both panels, the black line corresponds to $U = 0$, while lines of different colors distinguish different Stark field strengths. Other parameters, $N = 8$, $T = 0.1$,$\Delta=1$,$K_{tr}=50$.}
\label{DS_second_order}
\end{figure*}
\begin{figure*}[t]
\includegraphics[width=0.99\textwidth]{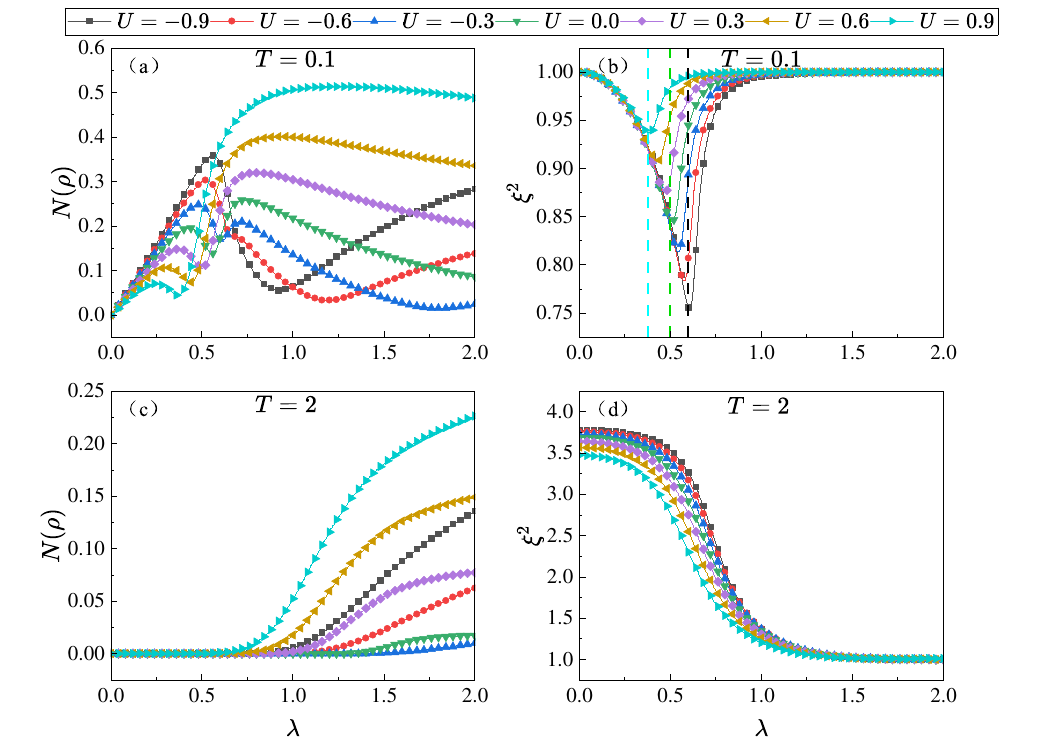}
\caption{The negativity $N(\rho )$ (panel (a) and (c)) and spin-squeezing parameter ${\xi^2}$ (panel (b) and (d)) as functions of coupling strength $\lambda$ under different Stark field strengths $U$, at temperature $T = 0.1$ (panel (a) and (b)) and $T = 2.0$ (panel (c) and (d)). In panel (a)-(d), lines of different colors correspond to different Stark field strengths: $U=-0.9, -0.6, -0.3, 0.0, 0.3, 0.6, 0.9$. Other parameters, 
$N = 8$ (panel (a) and (c)), $N = 32$ (panel (b) and (d)),
$\Delta=1$,$K_{tr}=50$.}
\label{DS_negativity}
\end{figure*}
\begin{figure*}
\includegraphics[width=0.99\textwidth]{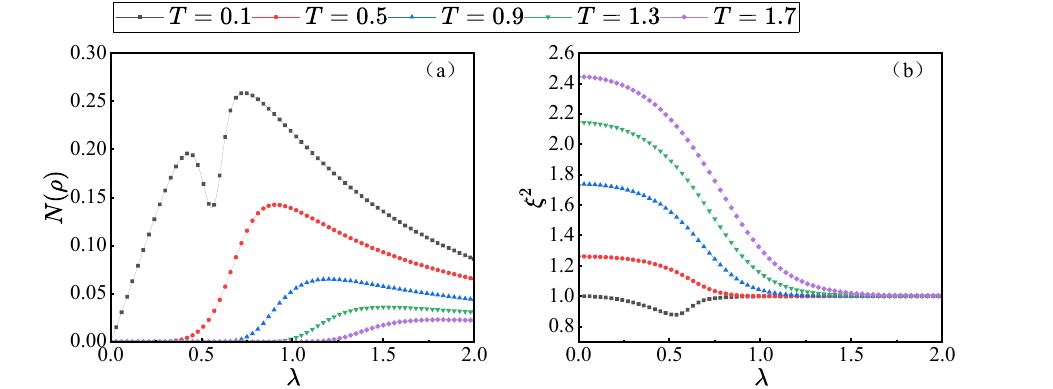}
\caption{
The negativity of $N(\rho )$ (panel (a)) and spin-squeezing parameter ${\xi^2}$ (panel (b)) as function of coupling strength $\lambda$ under different temperature $T$. In (a)-(b), lines of different colors correspond to different temperature: $T=0.1, 0.5, 0.9, 1.3, 1.7$. Other parameters, $U=0$, $N = 8$, $\Delta=1$, $K_{tr}=50$.}
\label{DS_T_negativity}
\end{figure*}

The traditional definition of the normalized zero-time delay two-photon correlation function is only applicable to the state where the optical field is weakly coupled to atoms~\cite{glauber1963quantum}. For calculations under arbitrary coupling strengths, it should be performed in the dressed states. Therefore, we discuss the generalized zero-time delay two-photon correlation function of the DS model, which is defined as
\cite{rabl2011photon}:
\begin{equation}
    {G^{(2)}}(0) = \frac{{({{({{\hat X}^ - })}^2}{{({{\hat X}^ + })}^2})}}{{{{(\hat X^ - {{\hat X}^ + })}^2}}},
\end{equation}
where ${\hat X^ + } =  - i\sum\limits_{k > j} {{\Delta _{kj}}} {\hat X_{jk}}|{\phi _j}\rangle \langle {\phi _k}|$, ${\Delta _{kj}}=\Delta _{k}-\Delta _{j}$ is the difference in energy levels, and ${\hat X_{jk}} = \langle {{\phi _j}|({{\hat a}^ \dagger} + \hat a)|{\phi _k}} \rangle$ describes the change from a high eigenstate to a low eigenstate. 

The two-photon correlation function quantifies the spatiotemporal correlation between photons. When \( g^{(2)}(0) < 1 \), the system exhibits anti-bunching, characterized by photons tending to be emitted separately, which is a typical quantum feature. On the other hand, when \( g^{(2)}(0) > 2 \), the system displays bunching effect, indicating that photons are highly clustered, meaning photons tend to be excited in clusters.

Although the traditional Dicke model provides a foundational framework for understanding quantum correlations in collective light-matter interactions and reveals rich two-photon correlation behaviors near the superradiant phase transition or spectral collapse critical points, these studies mainly focus on the cooperative effects induced by the coupling strength (\( \lambda \)) between the light field and atoms \cite{tiwari2023quantum}. The Dicke-Stark model studied in this paper introduces the Stark interaction term (\( \hat{H}_{\text{Stark}} \)), adding a crucial dimension to the standard Dicke Hamiltonian. The level shift introduced by this term provides a new independent physical parameter for manipulating photon statistical properties, such as \( g^{(2)}(0) \).

In this work, based on the DS model with $N = 8$, we investigate the variation of the zero-delay two-photon correlation function $G_8^{(2)}(0)$ in the thermodynamic equilibrium state with coupling strength $\lambda$ and Stark field strength $U$, 
as shown in Fig.~\ref{DS_second_order}, within the range of $\lambda \in [0, 1.2]$, $G_8^{(2)}(0)$ starts and ends at 2.0. With increasing $\lambda$, it first tends to a minimum value (less than 1), then rises to a maximum of approximately 50. This clearly depicts the evolutionary process of the system state: as the coupling strength increases, the state transitions from bunching to anti-bunching, then back to bunching. Furthermore, the bunching property gradually intensifies to an extreme, then returns to the same bunching property as in the weak coupling state. 

The Stark field strength affects both the magnitudes of the maximum and minimum values of $G_8^{(2)}(0)$ and the values of the coupling strength $\lambda$ corresponding to these maximum and minimum points. By comparing the curves in Figs.~\ref{DS_second_order}(a) and (b), it is found that when $U = -0.3$, the minimum value of $G_8^{(2)}(0)$ is the smallest and the maximum value is the largest.
For $(U > -0.3$ (with values sequentially set to -0.1, 0, 0.1, 0.3, 0.5, 0.7, 0.9, 1.0) or $U < -0.3$ (with values sequentially set to -0.5, -0.7, -0.9, -1.0), both the minimum and maximum values of $G_8^{(2)}(0)$  decrease successively. 

\subsubsection{\label{sec:negativity_squeezing}
Negativity and spin-squeezing}
To explore the quantum correlations between the light-matter constituents, we commence by calculating the quantum entanglement within bipartite systems. Among the various entanglement quantifiers available, we opt to use the negativity $N(\rho )$~\cite{eisert1999comparison,zyczkowski1998volume,plenio2005logarithmic,lee2000partial,vidal2002computable}
\begin{equation}~\label{sec:negativity}
    N(\rho ) = \frac{{||{\rho ^{{T_A}}}|{|_1} - 1}}{2},
\end{equation}
where ${\rho ^{{T_A}}}$ is the partial transpose of the quantum state $\rho $ with respect to atom subsystem and $\parallel {\rho ^{{T_A}}}{\parallel _1} = Tr|{\rho ^{{T_A}}}| = Tr\sqrt {{({\rho ^{{T_A}}})^\dagger }{\rho ^{{T_A}}}} $ is the trace norm or the sum of the singular value of the operator ${\rho ^{{T_A}}}$. Equivalently, the negativity can be computed as $N(\rho ) = 1/2\sum\limits_i {(|{\varepsilon _i}| - {\varepsilon _i})} $, where ${{\varepsilon _i}}$ are the eigenvalues of the partially transposed light-matter density matrix $\rho $. Note that $N(\rho ) = 0$ corresponds to separable (not entangled) quantum states, while $N(\rho)>0$ indicates that the quantum state is entangled. 

The spin-squeezing parameter is used to study atomic correlations and is also a measure of entanglement between atoms. According to the research by Kitagawa and Ueda~\cite{kitagawa1993squeezed}, the spin-squeezing parameter is defined as 
\begin{equation}
   {\xi ^2} = \frac{{2{{(\Delta {S_{{{\vec n}_ \bot }}})}^2}}}{J} = \frac{{4{{(\Delta {S_{{{\vec n}_ \bot }}})}^2}}}{N}, 
\end{equation}
where $\vec n \bot $ refers to an axis perpendicular to the mean spin $\langle \vec S\rangle $ and ${\Delta {S_{{{\vec n}_ \bot }}}} = \vec S \cdot {{\vec n}_ \bot }$. The spin squeezing parameter ${\xi ^2} < 1$ indicates that the system is spin squeezed.

In an open DS system with a fixed number of atoms $N=8,32$, the light-matter coupling strength $\lambda$, Stark field strength $U$, and temperature $T$ collectively influence quantum entanglement $N(\rho )$ and spin squeezing $\xi^2$, as shown in Fig.~\ref{DS_negativity}.

At a low temperature $T = 0.1$, within the range of coupling strength $\lambda \in [0, 2.0]$, negativity $N(\rho )  > 0$ and exhibits non-monotonic oscillatory behavior, as depicted in Fig.~\ref{DS_negativity}(a). This indicates that at low temperatures, entanglement persists in the thermal state of the system regardless of the strength of light-matter coupling.
At a high temperature $T = 2.0$, the oscillatory behavior of $N(\rho )$ vanishes. When the coupling strength is less than 0.5, $N(\rho )= 0$, which means the absence of entanglement in the state of the system. However, in regions with stronger coupling strengths for $\lambda >0.5$, $N(\rho ) > 0$  indicating the presence of quantum entanglement, as shown in Fig.~\ref{DS_negativity}(c).
The effect of $U$ on negativity $N(\rho )$ is notable. Within our calculated range $U\in [-0.9, 0.9]$, the maximum value of $N(\rho )$ occurs at $U = 0.9$, while the minimum value appears at $U = -0.3$. This phenomenon is consistent with the results regarding the influence of $U$ on $G_8^{2}(0)$ as presented in Fig.~\ref{DS_second_order}.

At $T = 0.1$, an obvious spin squeezing phenomenon is observed, as shown in Fig.~\ref{DS_negativity}(b). When the coupling strength $\lambda=0$, $\xi^2\sim 1$. As $\lambda$ gradually increases in the strong coupling region, $\xi^2$ decreases gradually, manifesting spin squeezing ($\xi^2 < 1$). $\xi^2$ reaches its minimum value near $\lambda \approx 0.5$.  Subsequently, with increasing $\lambda$, $\xi^2$ gradually rises and asymptotically approaches 1, indicating the disappearance of spin squeezing. 
Different values of $U$ not only modulate the squeezing strength but also shift the 
$\lambda$ position corresponding to maximum squeezing (minimum $\xi^2$).
It should be emphasized here that the $\lambda$ corresponding to the minimum value of $\xi^2$ is related to the SPT point. In Figure \ref{DS_negativity}(b), three vertical dashed lines mark the positions of the minimum values of $\xi^2$ when $U = -0.9, 0, 0.9$; 
The corresponding values of $\lambda $ agree well with the values of $\lambda_c=0.60, 0.50, 0.37 $ calculated according to Eq.(\ref{equ:lambda_c(T)}) at $N=32$.
At $T = 2$, as shown in Figure \ref{DS_negativity} (d), the spin squeezing vanishes. Within the range of coupling strength $\lambda\in [0, 2.0]$, $\xi^2$ monotonically decreases from approximately 2.5 and asymptotically approaches 1 with increasing $\lambda$.

Now, let $U = 0$ and take different temperatures $T = 0.1, 0.5, 0.9, 1.3, 1.7$ to study the influence of temperature on negativity $N(\rho)$ and spin squeezing $\xi^2$ in more detail. Fig.~\ref{DS_T_negativity}(a) shows that as $T$ increases, the oscillatory behavior of negativity $N(\rho)$ gradually disappears and its overall magnitude is suppressed. 
With increasing temperature, $N(\rho)$ approaches zero starting from smaller
values of $\lambda$: at $T = 0.5$, the negativity value is zero in the region $\lambda < 0.3$; at $T = 0.9$, the region where the negativity is zero expands to $\lambda < 0.7$; and at $T = 1.7$, this region expands further to $\lambda< 1.1$. This indicates that the degree of entanglement decreases with the increase of temperature, but the strong coupling between light and atoms is beneficial to the maintenance of entanglement.
Fig.~\ref{DS_T_negativity}(b) shows that an increase in temperature causes $\xi^2$ to increase rapidly. At $T = 0.5$, $\xi^2$ has already exceeded 1, indicating the disappearance of spin squeezing. This highlights the strong low-temperature dependence of spin squeezing in the open DS system.

\begin{figure*}[t]
\includegraphics{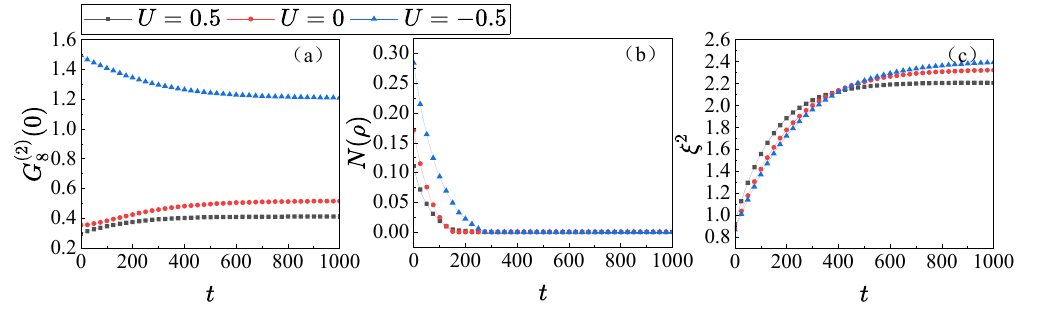}
\caption{Time evolution of (a) the zero-time delay two-photon correlation function $G^{(2)}(0)$, (b) the negativity $N(\rho)$, and (c) the spin squeezing parameter $\xi^2$. The black squares, red circles, and blue triangles correspond to Stark strengths $U = 0.5$, $U = 0$, and $U = -0.5$, respectively. Other simulation
parameters are $N = 8$, $\lambda = 0.5$, $\Delta = 1$, and $K_{tr} = 50$.}
\label{DK_T}
\end{figure*}

\subsection{\label{sec:properties2}Statistical properties of the thermodynamic non-equilibrium states }

Next, we briefly examine the time-evolution properties of the thermodynamic non-equilibrium states in the DS model governed by the dressed master equation Eq.~(\ref{eq:master_equation}), focusing on the effect of varying Stark interaction strengths \( U \). Specifically, we consider a DS system initially prepared in a low-temperature equilibrium state (\( T=0.1 \)) that is coupled to a high-temperature thermal bath (\( T=2.0 \)) at \( t=0 \). The initial and final states of the system are jointly determined by the parameters \( \lambda \), \( \Delta \), and \( U \). Figure \ref{DK_T} presents the time-evolution behaviors of the two-photon correlation function, Negativity, and spin squeezing under different Stark interaction strengths (\( U=-0.5, 0, 0.5 \)) with fixed \( \lambda = 0.5 \) and \( \Delta = 1 \).

Figure~\ref{DK_T}(a) illustrates the time evolution of the two-photon correlation function $G^{(2)}(0)$. For the initial system state at \( T=0.1 \) with \( U=-0.5 \), \( G^{(2)}(0) = 1.5 \), indicating that the system exhibits antibunching behavior ($G^{(2)}(0) > 1$). In the final system state at $T=2.0$, $G^{(2)}(0) = 1.2$, demonstrating that the system retains antibunching characteristics but with attenuated intensity. Similarly, at Stark strengths $U=0.5$ and $U=0$, $G^{(2)}(0) = 0.3$ and $G^{(2)}(0) = 0.4$ respectively, meaning that the system displays bunching behavior ($G^{(2)}(0) < 1$). In the final state at $T=2.0$, \( G^{(2)}(0) \) shifts to 0.4 and 0.5, respectively, such that the system remains in a bunching state with weakened intensity.
Further observation reveals that during the evolution of $G^{(2)}(0)$, for $U=-0.5$, when $t > 900$, $G^{(2)}(0)$ no longer shows significant variation, implying that the system has evolved into the thermal steady state corresponding to $T=2.0$. For $U=0$ and $U=0.5$, the system reaches the steady state at $T=2.0$ after $t > 500$ and $t > 600$, respectively. These results demonstrate that $U$ can modulate the evolution of $G^{(2)}(0)$, and a smaller value of $U$ leads to a longer time for $G^{(2)}(0)$ to reach its steady-state value.

Figure~\ref{DK_T}(b) depicts the dynamics of the negativity $N(\rho)$ as the system evolves from an initial low temperature $T=0.1$ towards a final thermal equilibrium at $T=2.0$. At the initial state ($T=0.1$), the negativity is approximately $0.28$ for $U = -0.5$, indicating a relatively high degree of entanglement. In contrast, the initial negativity values are significantly lower for \( U = 0 \) and \( U = 0.5 \), at approximately 0.15 and 0.08, respectively. As the system thermalizes to $T=2.0$, the negativity decays monotonically to zero for all three Stark interaction strengths. Interestingly, during the thermalization process, a crossover emerges between the negativity decay curves for $U = 0$ and $U = 0.5$, indicating a nontrivial role of $U$ in the entanglement dynamics. 
Nevertheless, in comparison with the systems at \( U = 0 \) and \( U = 0.5 \), the  \( U = -0.5 \) system maintains robust entanglement, persisting for a long duration and only disappearing when \( t > 400 \). Specifically, the system with $U=0$ loses its entanglement at $t \approx 200$, whereas the $U = 0.5$  system exhibits the most fragile entanglement, which vanishes as early as $t \approx 150$. These results confirm that $U = -0.5$ effectively extends the survival time of entanglement against thermal noise.

Figure~\ref{DK_T}(c) displays the evolution of the spin squeezing parameter $\xi^2$. Initially at $T=0.1$, the system exhibits spin squeezing ($\xi^2 < 1$) for the three Stark interaction strengths ($U=-0.5,0,0.5$). Specifically, for \( U = -0.5 \), the initial value of the squeezing parameter is \( \xi^2 \approx 0.85 \), indicating strong squeezing. In the case of \( U = 0 \), an intermediate squeezing effect is observed with \( \xi^2 \approx 0.90 \), while \( U = 0.5 \) exhibits the weakest initial squeezing, with \( \xi^2 \approx 0.95 \).
As time increases and the system heats up towards $T=2.0$, $\xi^2$ rises rapidly, leading to the vanishing of squeezing at different times. The squeezing effect disappears earliest for $U = 0.5$ at $t \approx 25$. For $U = 0$, the squeezing persists slightly longer, surpassing the standard quantum limit ($\xi^2 = 1$) at $t \approx 40$. In contrast, $U = -0.5$ maintains the squeezing state for the longest duration, vanishing at $t \approx 60$. Consequently, within the effective squeezing region ($\xi^2 < 1$), $U = -0.5$ demonstrates a clear advantage over both $U=0$ and $U=0.5$ by maintaining a lower $\xi^2$ value and a longer duration before the high temperature destroys the quantum correlations.

\section{CONCLUSION}
We examined the variation of the ground-state average photon number with coupling strength and its time evolution, demonstrating the quantum phase transition phenomenon and the shift of the critical point with Stark field strength. At N=128, a clear critical point becomes observable. Within the range of Stark strengths $U\in [-1.5,1.5]$,  the critical point shifts towards smaller coupling strengths as $U$ increases, which is consistent with the theoretical calculation results.

As the coupling strength increases, the two-photon correlation function \(G^{(2)}(0)\) indicates that the light field evolves from a thermal state to anti-bunching behavior, then to re-entrant bunching, and finally back to a thermal state. Crucially, the Stark field strength modulates both the extremal values (maximum and minimum) of \(G^{(2)}(0)\) and the corresponding coupling strengths \(\lambda\).

The light-atom coupling strength $\lambda$, Stark strength $U$, and temperature $T$ collectively influence quantum entanglement $N(\rho )$ and spin squeezing $\xi^2$.
At low temperatures (\(T = 0.1\)), the negativity $N > 0$, indicating that the system is in an entangled state. However, as the temperature increases, the entanglement of the system state begins to vanish at small coupling strengths. At \(T = 2.0\), the system state remains entangled only when the coupling strength is sufficiently large. Thus, strong coupling is conducive to maintaining the entanglement of the system state. 
Atomic squeezing is sensitive to temperature. At \(T = 0.1\), atomic squeezing can be observed in the region where \(\lambda\) is small, and an increase in the negative Stark field strength enhances the squeezing property with the minimum of atomic squeezing appearing near the SPT. However, atomic squeezing rapidly vanishes as the temperature increases.

While thermal noise eventually breaks down quantum states, the Stark interaction regulates the speed of this process. We find that negative interactions ($U=-0.5$) help protect the system. Under these conditions, the negativity $N(\rho )$ and spin squeezing $\xi^2$ parameter increases more slowly than in the positive interaction($U=0.5$) case, indicating that negative interactions($U=-0.5$) make the quantum resources more robust against thermal noise.

Based on the systematic study of the Dicke-Stark model presented in this work, future experimental inquiries using platforms such as superconducting circuits and trapped ions~\cite{baksic2014controlling,forn2019ultrastrong,kim2011quantum,schuckert2025observation} could deeply explore nonequilibrium dynamical behaviors. For instance, quantum quenching protocols could be employed to investigate dynamic phase transitions and the evolution of photon statistics. Theoretically, extending the system to larger scales ($N \gg 128$) would facilitate the analysis of finite-size effects and the scaling laws governing quantum correlations. It would also be valuable to apply the coherent-state method to more complex light-matter coupling models—such as those incorporating $A^2$ terms or atomic interactions—to verify its universality. Finally, exploring the deep strong coupling regime ($\lambda/\omega_c \sim 1$) or regimes with larger Stark shifts ($U/\omega_c > 0.2$) may reveal exotic phases of matter, such as superradiant glass phases, thereby opening new avenues for quantum sensing and the generation of non-classical light sources.

\section*{Acknowledgments}
We acknowledge useful discussions with Jiasen Jin, Qing-Hu Chen and He-Guang Xu. This work is supported by the Science and Technology Projects of China Southern Power Grid (YNKJXM20220050)

\appendix
\section{\label{FA} The extended coherent state approach}

An extended coherent bosonic state approach is proposed to accurately calculate the energy spectrum and eigenstates of the finite-size Dicke model with a small photon number cutoff~\cite{chen2008numerically}. This method is also applicable to the finite-size DS model. Before the introduction of the extended coherent bosonic state approach, we rotate the collective spin operators with $\pi /2$ along $\hat{J}_{y}$ by ${\hat H} = \exp(i\pi {\hat J_y}/2){\hat H_{DS}}\exp( - i\pi {\hat J_y}/2)$, resulting in
\begin{equation}~\label{H}
\hat H = \omega {\hat a^\dagger }\hat a - (\frac{\Delta }{2} + \frac{U}{{2N}}{\hat a^\dagger }\hat a)({\hat J_ + } + {\hat J_ - }) + \frac{{2\lambda }}{{\sqrt N }}({\hat a^\dagger } + \hat a){\hat J_z}.
\end{equation}
in which ${\hat { J } _ { \pm } = \hat {J} _ {x} \pm  \hat {J} _ {y} }$ are the raising and lowering operators of the atomic states $\left| {j,m} \right\rangle$, $m =-j, -j+1, \ldots,j-1, j $ and $j = N/2$. ${\hat J_ \pm }|j, m\rangle  = j_m^ \pm |j,m \pm 1\rangle $, with $j_m^ +  = \sqrt {j(j + 1) - m(m \pm 1)} $. 
The states of the total system in Hilbert space can be expressed in terms of the direct product basis of the bosonic field states and the atomic states,$\left\{ {|{\varphi}{ \rangle _b} \otimes |j,m\rangle } \right\}$.
$|{\varphi}{\rangle _b}$ is bosonic field state, which can be expanded in different spaces, such as the Fock state space
$$
|{\varphi}{\rangle _b} = \sum\limits_{n }  {C_n}|n{\rangle}.
$$
Considering a displacement operator $\hat{D}(g_m)=\exp(g_m\hat{a}-g_m\hat{a}^{\dagger})$, with the ${g_m} = 2\lambda m/\omega \sqrt N$, the displaced operator ${\hat{A}_m} =\hat{D}^{\dagger} \hat{a}\hat{D}=\hat{a} + {g_m}$, ${\hat{A}^{\dagger}_m} =\hat{D}^{\dagger} \hat{a}^{\dagger}\hat{D} =\hat{a}^{\dagger} + {g_m}$ serve as the creation and annihilation operators of the extended coherent state space, which is defined as:
$$|k{\rangle _{A_m}={\frac{1}{{\sqrt {k!} }}}(\hat A_m^ \dagger  )^k}|0{\rangle _{A_m}},$$ with $|0{\rangle} _{A_m}=\hat{D}^ \dagger|0\rangle$ is the vacuum state in the extended coherent state space. Then, the bosonic field state can also be expanded in the extended coherent state space as
$$
|{\varphi}{\rangle _b} = \sum\limits_{k = 0}^{K_{tr}} {C_{k}}|k{\rangle _{A_m}}.
$$
Obviously, the expansion of the photonic state in the extended coherent state space will include all Fock states, thus a relatively small truncation number $K_{tr}$ can yield accurate calculation results. 

In the direct product space of the generalized coherent state space and the atomic state space, the eigenstate determined by $\hat{H}|\psi_n\rangle=E_n|\psi_n\rangle$  is assumed to be:
\begin{equation}~\label{equ:state}
|{\psi_n}{\rangle} = \sum\limits_{m,k} {C^n_{m,k}}|j,m\rangle|k{\rangle _{A_m}}.
\end{equation} 
Then, we arrive at the equation for $E_n$:
\begin{widetext}
\begin{equation}
\begin{array}{l}
\sum\limits_{m,k} {\omega C_{m,k}^n} (k - g_m^2)|j,m\rangle|k{\rangle_{{A_m}}}\\
 - \sum\limits_{m,k} {C_{m,k}^n} \left[ {\frac{\Delta }{2} + \frac{U}{{2N}}(k + g_m^2)} \right]\left(j_m^ + |j,m + 1\rangle|k{\rangle _{{A_m}}}
 + j_m^ - |j,m - 1\rangle |k{\rangle _{{A_m}}}\right)\\
 + \frac{U}{{2N}}\sum\limits_{m,k} {C_{m,k}^n} {g_m}(\sqrt {k + 1} j_m^ + |j,m + 1\rangle|k + 1{\rangle _{{A_m}}}+ \sqrt k j_m^ + |j,m + 1\rangle |k - 1{\rangle _{{A_m}}})\\
 + \frac{U}{{2N}}\sum\limits_{m,k} {C_{m,k}^n} {g_m}(\sqrt {k + 1} j_m^ - |j,m - 1\rangle |k + 1{\rangle _{{A_m}}}+ \sqrt k j_m^ - |j,m - 1\rangle |k - 1{\rangle _{{A_m}}})\\
 = {E_n}\sum\limits_{m,k} {C_{m,k}^n} |j,m|k{\rangle _{{A_m}}}.
 \end{array}
\end{equation}
\end{widetext}
Left multiply $\left\{ {\left\langle l \right|\left\langle {n,j} \right|} \right\}$ with $n =  - j, - j + 1, \ldots ,j$. We obtain the Eq. \ref{equ:E_n} by the expansion coefficients $C^n_{m,k}$.

\section{\label{FB} Error analysis of DCS and DFS method in the DS Model}
In this section, we will systematically verify the applicability and convergence of DCS and DFS method in the DS model.

\begin{figure*}
\includegraphics{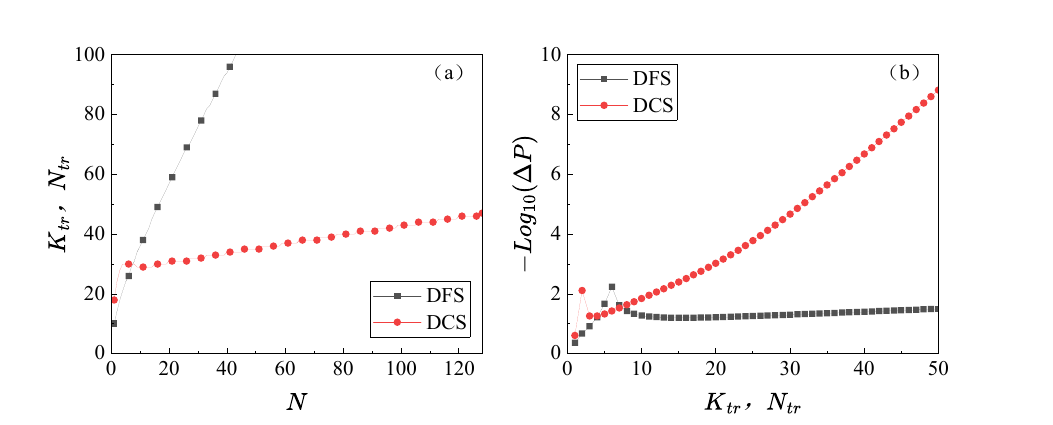}
\caption{Comparison of the ground-state calculation accuracy between the DCS (red lines) and DFS (black lines) method. (a). Energy error analysis: \( K_{tr}\) and \(N_{tr} \) represent the minimum truncation number required to achieve a calculation accuracy of \( 10^{-6} \) for atom number \( 1\le N\le 128 \). 
(b). Wavefunction error analysis: 
Effects of the photon number truncations $N_{tr}$ and $K_{tr}$ on calculation of the error $-\log_{10}(\Delta P)$ of the ground-state wavefunction with atom number $N=128$.
The vertical axis represents calculation accuracy, with higher values of \(-\log_{10}(\Delta P) \), indicating smaller wavefunction errors. Other parameters: \( U = 1.0 \), \( \Delta = 1 \), \( \lambda = 0.5 \).}
\label{Ground state convergence}
\end{figure*}

\begin{figure*}
\includegraphics{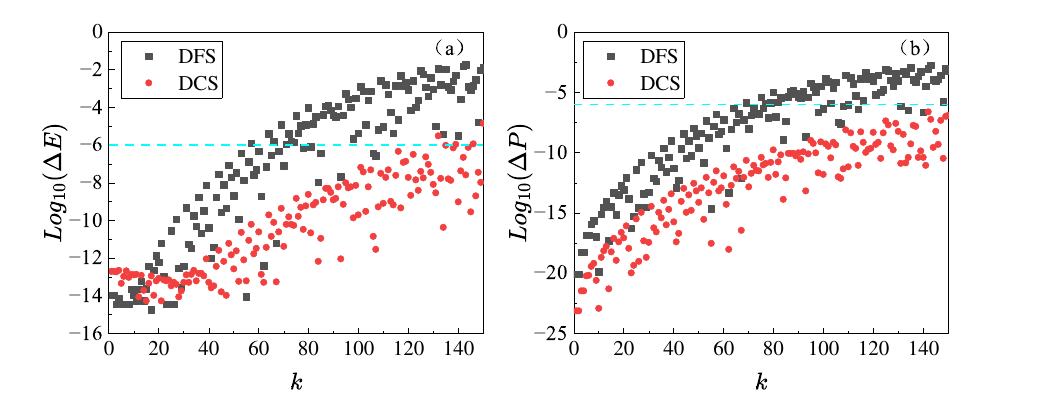}
\caption{Comparison of the accuracy between DCS and DFS method in the excited-state calculation of the Dicke model. (a) Excited-state energy error \(\log_{10}(\Delta E) \) as a function of the number of excited state \( k \). (b) Excited-state wavefunction error \(\log_{10}(\Delta P) \) as a function of the excited state \( k \). 
Red circles represent the results of the DCS method, while black squares represent the results of DFS method. Blue dashed lines serve as a reference of calculation accuracy $10^{-6}$. Other parameters: \( N = 32 \), \( U = 1.0 \), \( \lambda = 0.5 \), \( \Delta = 1 \), and \(N_{{tr}}= K_{{tr}} = 50 \).}
\label{Excited state convergence}
\end{figure*}

First, we discuss the error of the ground state. Based on the error estimation method described in Ref.\cite{bastarrachea2014efficient}, \[\Delta E = \left| {E({K_{tr}} + 1) - E(K_{tr})} \right|,\]
$E(K_{tr})$ represents the energy magnitude under the truncation of photon numbers $K_{tr}$.
We calculated the minimum photon number truncations, \(N_{tr}\) for the DFS method and \(K_{tr}\) for the DCS method, required to achieve a ground-state energy error below \(10^{-6}\) with parameters \(u = 1\), coupling strength \(g = 1\) and the maximum number of atoms \(N = 128\). As shown in Fig.~\ref{Ground state convergence}(a), for a small number of atoms ($N < 8$), the DFS method requires a smaller photon number truncation than that the DCS method requires, i.e., \(N_{tr}>K_{tr}\). 
In contrast, as $N$ increases, $N_{tr}$ grows rapidly, whereas $K_{tr}$ remains significantly smaller. At $N = 128$, $K_{tr}$ does not exceed 50. 
Furthermore, based on the wavefunction error analysis in \cite{hirsch2014efficient,bastarrachea2014fidelity,bastarrachea2014peres},
An error parameter describing the calculation accuracy of wavefunction is defined as follows:
\[\Delta P{\rm{ }} = \sum\limits_{m =  - j}^j {{{\left| {{C_{m,{K_{tr}+1}}}} \right|}^2}} ,\]
where $C_{m,{K_{tr}+1}}$ are the expansion coefficients in Eq.(\ref{equ:state}), and can be determined by solving Eq. (\ref{equ:E_n}). 
Fig.~\ref{Ground state convergence}(b)
presents the effects of the photon number truncations $N_{tr}$ and $K_{tr}$ on the ground-state wavefunction error $\Delta P$ for a system with atom number
$N=128$.
Overall, increasing the truncations $N_{tr}$ and $K_{tr}$ reduce the calculation error of the wavefunction. Notably, the error $\Delta P$ decreases rapidly as $K_{tr}$ increases, falling below $10^{-9}$ at $K_{tr}=50$. Conversely, the wavefunction error \(\Delta P \) shows little sensitivity to $N_{tr}$, remaining above $10^{-2}$ for $N_{tr}<50$.
Therefore, Fig.~\ref{Ground state convergence} (a) and (b) demonstrate the high precision of DCS approach in describing the ground state.

Second, we discuss the error of the excited states.
In Fig.~\ref{Excited state convergence},
we present energy error \( \log_{10}(\Delta E) \) and  wavefunction error \( \log_{10}(\Delta P) \)  for the first 150 eigenstates with \(N = 32\) and the truncations $N_{tr}=K_{tr}=50$. Figs.~\ref{Excited state convergence}(a) and ~\ref{Excited state convergence}(b) show the energy and wavefunction errors for these 150 energy levels, respectively. 
It can be observed that the errors of energy levels and wavefunctions calculated using the DCS method are significantly lower than those obtained via the DFS method. When \(K = 50\), the calculation errors of excited-state energy levels and wavefunctions using the DCS method are well controlled below \(10^{-6}\).


Here, it is necessary  to confirm that selecting 150 energy levels is sufficient to capture the physical properties of the system. Using Eq.(\ref{equ:density}), we calculated the sum of probabilities of these energy levels under the thermodynamic equilibrium distribution at a temperature of \(T = 2\). The results show that the sum of probabilities exceeds 0.9958, indicating that the first 150 energy levels almost completely describe the state of the system.

\section{\label{FC} Theoretical solution of  SPT in infinitely sized Dicke model}
\subsection{\label{FC1} SPT in isolated infinite-size DS system}
In the infinite-size DS model, the system possesses collective symmetry, which can give rise to the occurrence of SPT. For an isolated DS system, the SPT of the system can be solved analytically.
We employ the Holstein-Primakoff transformation to express the collective spin operators in terms of bosonic operators:
${{\hat J}_ + } = {{\hat b}^ \dagger }\sqrt {N - {{\hat b}^ \dagger }\hat b} $,
${{\hat J}_ - } = \sqrt {N - {{\hat b}^ \dagger }\hat b} \hat b$ and
${{\hat J}_z} = {{\hat b}^ \dagger }\hat b - \frac{N}{2}$.
Here, the bosonic operators satisfy the canonical commutation relation
$\left[ {\hat b,{{\hat b}^ \dagger }} \right] = 1$.
The bosonic modes are then displaced relative to their ground-state expectation values as:
$\hat b \to \beta  + \hat d$~\cite{zhai2025stark,kirton2019introduction,chen2024phase}.
Where, the spin fluctuations obey with 
$\left[ {\hat d,{{\hat d}^ \dagger }} \right] = 1$ and $\beta  = \langle GS|\hat b|GS\rangle$, with $|GS\rangle $ is the ground state. To the zeroth-order approximation, the system Hamiltonian expressed by Equation (\ref{H_D}) transforms into:
\begin{equation}\label{equ:H_A}
    {\hat H_{A}} = \omega _c^\prime {{\hat A}^ \dagger}\hat A - \frac{{4g^2\beta {^2}}}{{\omega _c^\prime }} + \Delta {\beta ^2} - \frac{{\Delta N}}{2},
\end{equation}
in which, ${{\hat A}^\dagger}={{\hat a}^\dagger}+\frac{{2g\beta }}{{\omega _c^\prime }}$ with $\omega _c^\prime=\omega+U{\beta ^2}/N - U/2$. In the coherent state space, $\hat{H}_A$ is a diagonal matrix. Then, the energy of the photon vacuum state can be rewritten as: 
$$\frac { E _ { G } } { N } = \Delta \left( \phi - \frac { 1 } { 2 } \right) - \frac { 4 g ^ { 2 } \phi ^ { 2 } ( 1 - \phi ) } { \omega + U \left( \phi - \frac { 1 } { 2 } \right) },$$
where $ \phi = \beta ^ { 2 } / N$. The symmetry-breaking condition of the symmetric phase corresponds to the zero point of the second derivative of $ E _ { G }$ at $\phi = 0$, thereby deriving the critical coupling strength:
\begin{equation}
    \lambda_c = \frac{1}{2}\sqrt{\Delta\left(\omega - \frac{U}{2}\right)}.
\end{equation}


\subsection{ \label{FC2}SPT in finite-temperature infinite-size DS system} 
The equilibrium state of a system at finite temperatures is determined by the minimization of the free energy, denoted as $F = - \frac{1}{\beta }\ln Z,\beta  = \frac{1}{{{k_B}T}}$, where $Z$ is the partition function.
By employing the mean-field approximation, neglecting photon-atomic field coupling, and applying the displacement transformation $\hat a = \tilde{\hat a} + \sqrt N \alpha$, the system Hamiltonian decomposes into $\hat H_{ph}$ and $\hat H_{at}$, describing the photon field and atomic states, respectively.
\begin{equation}
\begin{array}{l}
{\hat {H}_{ph}} = \omega ({{\tilde{\hat a}}^\dagger }\tilde {\hat a} + \sqrt N \alpha (\tilde {\hat a} + {{\tilde {\hat a}}^\dagger }) + N{\alpha ^2}),\\
{\hat H_{at}} = (\Delta  + U{\alpha ^2}){J_z} + 4\lambda \alpha {\hat J_x}.
\end{array}
\end{equation}
The total partition function is approximated as the product of the photon part and the atomic part $Z \approx {Z_{ph}} \cdot {Z_{at}}(\alpha )$.
The photon part of the partition function is independent of the displacement parameter $\alpha$:
\begin{equation}
\begin{array}{c}
{Z_{ph}} = {\rm Tr}{e^{ - \beta \omega ({{\tilde{\hat a}}^\dagger } + \sqrt N \alpha )(\tilde {\hat a} + \sqrt N \alpha )}} \\= \sum\limits_{n = 0}^\infty  {{e^{ - \beta \omega n}}}  = \frac{1}{{1 - {e^{ - \beta \omega }}}}.
\end{array}
\end{equation}
therefore, we have ${F_{ph}} = \frac{1}{\beta }\ln (1 - {e^{ - \beta \omega }})$.
Assuming the atoms constitute independent spin-$\frac{1}{2}$ systems, with the Hamiltonian for each atom given by: 
${h^{(1)}} = \frac{{\Delta  + U{\alpha ^2}}}{2}{\sigma ^z} + 2\lambda \alpha {\sigma ^x},$ 
the eigenvalues can be solved to be ${E_ \pm } =  \pm \frac{1}{2}\sqrt {{{(\Delta  + U{\alpha ^2})}^2} + {{(4\lambda \alpha )}^2}}$,
and the atomic part of the partition function is
\begin{equation}{Z_{at}}(\alpha ) = {\left[ {2\cosh (\beta \frac{{\phi (\alpha )}}{2})} \right]^N},
\end{equation}
where, $\phi (\alpha ) = \sqrt {{{(\Delta  + U{\alpha ^2})}^2} + {{(4\lambda\alpha )}^2}}$. 
Then, the total free energy density is
\begin{equation}
f(\alpha ) = F(\alpha )/N=\omega {\alpha ^2} - \frac{1}{\beta }\ln \left[ {2\cosh\left( {\frac{\beta }{2}\phi (\alpha )} \right)} \right].
\end{equation}
Computing the second derivative of $F(\alpha )$ in $\alpha = 0$, the symmetry-breaking condition for the symmetric phase corresponds to the zero of the second derivative. thereby, yielding the critical coupling condition of SPT in n open DS system:
\begin{equation}
\lambda_c(T) = \frac{1}{2} \sqrt{\Delta \left[ \frac{\omega}{\tanh\left(\frac{\Delta}{2k_B T}\right)} - \frac{U}{2} \right]}
.
\end{equation}
This is a temperature-dependent critical condition, which is helpful for discussing the properties of the DS model in quantum open systems, as detailed in Section~\ref{sec:negativity_squeezing}. When $T = 0$, $\lambda_c(T)$ reverts to the result of Eq.~(\ref{equ:lambda_c}) for the isolated DS system.


\bibliography{DSstark}

\end{document}